\documentclass[aps,reprint,onecolumn]{revtex4-2}

\usepackage{amsmath,amssymb,bm,graphicx}

\begin{document}

\title{Vibrational Mpemba relaxation in a linear damped elastic system}

\author{Kiwamu Yoshii}
\affiliation{Department of Applied Physics, Tokyo University of Science, 6-3-1 Niijuku, Katsushika-ku, Tokyo 125-8585, Japan}

\author{Satoshi Takada}
\email{takada@go.tuat.ac.jp}
\affiliation{Department of Mechanical Systems Engineering, Tokyo University of Agriculture and Technology, 2-24-16 Naka-cho, Koganei, Tokyo 184-8588, Japan}

\begin{abstract}
We show that a linear damped elastic system can display a Mpemba-like reversal of vibrational relaxation without any nonlinear constitutive response or amplitude-dependent damping.  
The mechanism is purely modal.
For a one-dimensional Kelvin--Voigt elastic medium, the decay rate of the $n$-th vibrational mode scales as $\gamma_n\propto n^2$.  
We formulate relaxation in terms of a positive modal envelope energy, thereby excluding crossings caused merely by oscillation phase.  
A one-parameter family of initial states is then constructed such that increasing the initial elastic excitation simultaneously reduces the projection onto the slow fundamental mode.  
In the invariant two-mode subspace, the relaxation-order crossing is obtained analytically and, for this family, occurs at a universal dimensionless time independent of the selected pair of initial states.  
In the limiting case where the slowest mode is absent, the dominant relaxation rate changes discontinuously, yielding a direct vibrational analogue of the strong Mpemba effect.
We further show that the same relaxation-order reversal occurs in the ordinary mechanical vibrational energy, which decreases monotonically despite retaining the oscillatory dynamics of the underlying modes.
\end{abstract}

\maketitle

\section{Introduction}
The Mpemba effect, originally discussed by Mpemba and Osborne in the context of the anomalous freezing of water~\cite{Mpemba1969}, has come to denote a much broader class of anomalous relaxation processes in which a state initially farther from a stationary state overtakes one that starts closer.
Although the interpretation of the original water experiments remains subtle~\cite{Burridge2016}, a spectral viewpoint has provided a general framework for understanding such relaxation reversals~\cite{LuRaz2017}.
In particular, a reduced or vanishing projection onto the slowest relaxation mode underlies the strong Mpemba effect~\cite{Klich2019}, which has also been demonstrated experimentally in a colloidal system~\cite{Kumar2020}.
Related studies have connected anomalous relaxation to metastability and nonequilibrium free-energy landscapes~\cite{Chetrite2021}, while recent experiments have emphasized that heating and cooling far from equilibrium can follow intrinsically different relaxation pathways~\cite{Ibanez2024}.
Recent reviews summarize the rapidly expanding range of classical and quantum realizations and the broader theoretical framework of anomalous relaxation speedups~\cite{Bechhoefer2021,AresReview2025,TezaReview2026}.

Classical nonequilibrium systems provide several particularly transparent examples.
Mpemba relaxation has been studied in driven and freely cooling granular gases~\cite{Lasanta2017}, in exactly tractable granular Maxwell models~\cite{Biswas2020}, and in sheared inertial suspensions~\cite{Takada2021}.
The dependence of the phenomenon on the choice of distance from the stationary state has also been examined explicitly~\cite{Biswas2023}.
More recently, extensions to granular gases with velocity-dependent restitution have revealed crossings in both thermal and rheological observables~\cite{Kikuchi2026}.
An exactly solvable two-dimensional bistable model further illustrates how the initial-state dependence of a slow spectral amplitude can generate a crossing in a positive global distance from equilibrium~\cite{HayakawaTakada2026}.
These examples emphasize that the Mpemba effect is not tied to temperature itself, but more generally to the redistribution of an initial state among relaxation channels with different decay rates.

Quantum counterparts exhibit a similarly broad spectral structure.
Accelerated relaxation has been demonstrated in Markovian open quantum systems by suppressing the slowest Liouvillian mode~\cite{Carollo2021}, in quantum dots coupled to reservoirs~\cite{Chatterjee2023}, and in non-Hermitian relaxation involving exceptional points and oscillatory modes, where multiple crossings can occur~\cite{Chatterjee2024}.
Quantum Mpemba effects have also been connected to symmetry restoration in isolated many-body systems~\cite{Ares2023,Rylands2024} and observed in trapped-ion quantum simulators~\cite{Joshi2024,Aharony2024,Zhang2025}.
Together, these developments reinforce the view that anomalous relaxation is fundamentally associated with the modal or spectral composition of the initial state rather than with a particular microscopic mechanism.

Of particular relevance to the present work, Mpemba-like behavior has previously been reported in vibrating mechanical systems.
Greaney \textit{et al.} observed an anomalously rapid ring-down of strongly excited flexural modes in molecular-dynamics simulations of carbon-nanotube resonators~\cite{Greaney2011}.
In that system, however, a simple linear damping description was insufficient: the anomalous relaxation was attributed to nonlinear dissipation together with an internal degree of freedom associated with a heterogeneous population of background phonon modes.
More recently, the Mpemba effect has been shown to arise within linear-response many-body dynamics through the spectral geometry of the relaxation operator~\cite{BenAbdallah2026}.
Related work on linearly coupled oscillators has demonstrated Mpemba crossings in the presence of parametric driving and thermal noise~\cite{Pahlevani2026}, while the Mpemba effect has also been extended explicitly to underdamped Brownian dynamics~\cite{Shapira2026}.
These results establish that neither inertia nor linear dynamics by itself precludes anomalous relaxation.
They nevertheless leave open a particularly simple mechanical question: can a relaxation-order reversal arise in a passive deterministic elastic continuum whose normal modes remain exactly decoupled?

This question is naturally connected with classical vibration theory.
Linear viscoelastic constitutive models, including the Kelvin--Voigt model, provide standard descriptions of internal material dissipation and vibration damping in solids and structures~\cite{Christensen1982,Nashif1985,Lakes2009,Zhou2016,Shu2022}.
A central result of linear vibration theory is that certain damping operators permit a classical modal representation in which the damped equations are simultaneously diagonalized and the normal coordinates evolve independently~\cite{Caughey1960,CaugheyOKelly1965,Woodhouse1998,Adhikari2006}.
For Kelvin--Voigt-type internal damping, the dissipative contribution contains spatial derivatives and therefore produces a decay rate that depends on wavelength or mode number.
The decay and stability properties associated with Kelvin--Voigt damping have consequently been studied extensively for strings, beams, and elastic wave equations~\cite{LiuLiu1998,LiuLiu2002,BurqSun2022}.
This familiar mechanical setting provides an unusually clean platform for asking whether mode-dependent damping alone is sufficient to generate Mpemba-type relaxation.

Here we answer this question affirmatively for a one-dimensional Kelvin--Voigt viscoelastic medium.
Because the elastic and viscous terms contain the same spatial operator, the normal modes are simultaneously diagonalized and evolve independently.
Nevertheless, their decay rates are strongly mode dependent, $\gamma_n\propto n^2$.
By preparing two initial states with different modal compositions, an initially more strongly excited state can therefore contain a smaller projection onto the slowly relaxing fundamental mode and subsequently overtake a less excited state.
To distinguish this relaxation reversal from trivial crossings generated by oscillation phase, we introduce a positive modal envelope energy and restrict the analysis to an underdamped invariant modal subspace.
Within an exact two-mode subspace, the crossing time is obtained analytically and, for the family considered below, is independent of the particular pair of initial states.
We further show that the same relaxation-order reversal occurs in the ordinary mechanical vibrational energy, which decreases monotonically in time.
When the slowest mode is removed completely, the dominant long-time relaxation rate switches to that of a faster mode, providing a direct vibrational analogue of the strong Mpemba effect.
Thus, in contrast to the earlier carbon-nanotube scenario, neither nonlinear dissipation, mode coupling, stochastic forcing, nor amplitude-dependent damping is required.

The remainder of this paper is organized as follows.
Section~\ref{sec:model} introduces the Kelvin--Voigt model and the modal relaxation measures.
Section~\ref{sec:Mpemba} derives the Mpemba crossing in an exact two-mode invariant subspace and demonstrates the corresponding reversal in the ordinary mechanical vibrational energy.
Section~\ref{sec:strong_Mpemba} discusses the strong vibrational Mpemba limit in which the slowest mode is eliminated.
Finally, Sec.~\ref{sec:conclusion} summarizes the mechanism and discusses its extension to more general linear mechanical systems.
Appendix~\ref{sec:derivation} provides a derivation of the positive modal envelope energy and its exponential decay law for underdamped modes.

\section{Model and modal relaxation}\label{sec:model}
\begin{figure}
    \centering
    \includegraphics[width=0.5\linewidth]{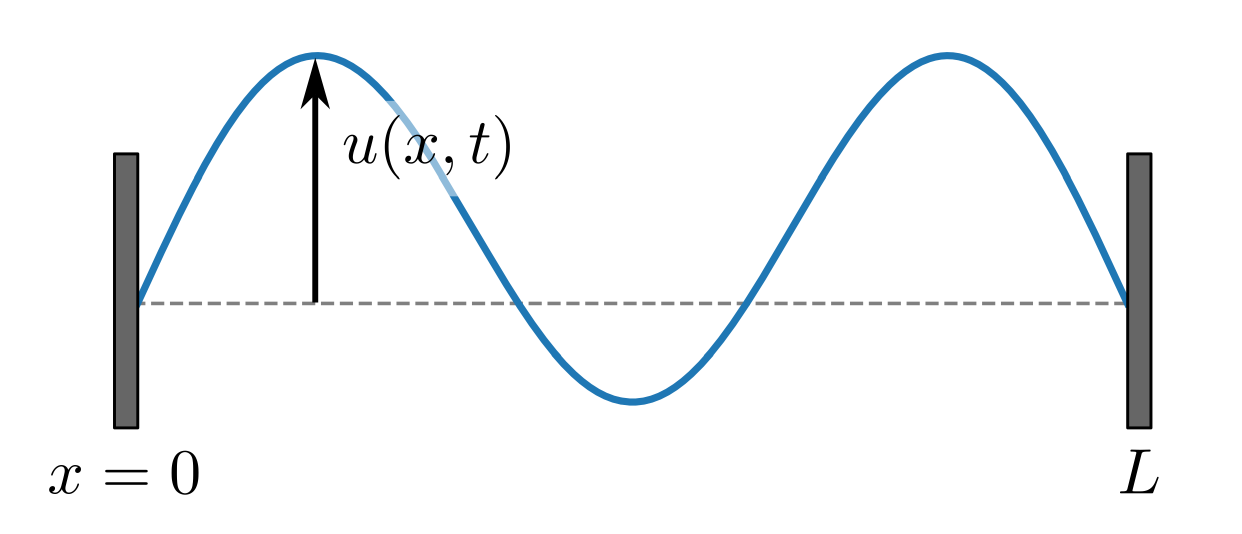}
    \caption{Schematic of the system.
    Fixed-end elastic systems is released at $t=0$.}
    \label{fig:setup}
\end{figure}

As illustrated in Fig.~\ref{fig:setup}, we consider a one-dimensional Kelvin--Voigt-type linear viscoelastic layer~\cite{Christensen1982,Nashif1985,Lakes2009,Zhou2016} of length $L$ ($0\le x\le L$), and denote the scalar shear displacement at each point by $u(x,t)$.
The displacement $u(x,t)$ evolves according to
\begin{equation}
    \rho u_{tt}
    = G u_{xx}
    + \eta u_{txx},
    \label{eq:pde}
\end{equation}
where $\rho$ is the mass density, $G$ is the shear modulus, and $\eta$ is the Kelvin--Voigt viscosity.
Hereafter, subscripts denote partial derivatives with respect to the corresponding variables~\cite{LiuLiu2002,BurqSun2022}.
We assume fixed-end boundary conditions, $u(0,t)=u(L,t)=0$.

Under the fixed-end conditions, the displacement can be expanded as~\cite{Achenbach1973,Ewins2000}
\begin{equation}
    u(x,t)=\sum_{n=1}^{\infty}q_n(t)\phi_n(x),
    \quad
    \phi_n(x)=\sqrt{\frac{2}{L}}\sin (k_nx),
    \label{eq:u_expansion}
\end{equation}
where $k_n=n\pi/L$.
Substituting Eq.~\eqref{eq:u_expansion} into Eq.~\eqref{eq:pde}, we find that $q_n(t)$ satisfies
\begin{equation}
    \ddot q_n
    +2\gamma_n\dot q_n
    +\omega_n^2q_n=0.
    \label{eq:mode}
\end{equation}
The equations are independent for each mode $n$, with
\begin{equation}
    \omega_n := \sqrt{\frac{G}{\rho}}k_n,\quad
    \gamma_n := \frac{\eta k_n^2}{2\rho}
    = n^2\gamma_1,
\end{equation}
and
\begin{equation}
    \gamma_1 := \frac{\eta \pi^2}{2\rho L^2}.
    \label{eq:gamma1}
\end{equation}

The exact modal decoupling in Eqs.~\eqref{eq:mode}--\eqref{eq:gamma1} is a continuum analogue of classical, or proportional, damping, for which the damping operator is diagonalized by the same modal basis as the undamped elastic operator~\cite{Caughey1960,CaugheyOKelly1965,Woodhouse1998,Adhikari2006}.
In the present Kelvin--Voigt model, this property follows directly from the fact that the elastic and viscous terms contain the same spatial operator $\partial_x^2$.

Instantaneous displacement or stress is not an appropriate measure of relaxation for the present purpose, because crossings can occur solely as a consequence of phase differences.
Indeed, oscillatory relaxation associated with complex eigenvalues is known to produce multiple Mpemba crossings in dissipative systems~\cite{Chatterjee2024}.
In the present study, by contrast, we remove the rapid phase oscillation and isolate the decay of the vibrational envelope.

For this purpose, we write the displacement of each mode as
\begin{equation}
    q_n(t)=\mathrm{e}^{-\gamma_n t}y_n(t),
    \label{eq:transform}
\end{equation}
and introduce the positive modal envelope energy
\begin{equation}
    \mathcal{E}_n(t)
    =
    \frac{\rho}{2}
    \left[
        (\dot q_n+\gamma_n q_n)^2
        +\Omega_n^2q_n^2
    \right].
    \label{eq:envenergy}
\end{equation}
A detailed derivation is given in Appendix~\ref{sec:derivation}.
This quantity satisfies
\begin{equation}
    \mathcal{E}_n(t)
    =
    \mathcal{E}_n(0)e^{-2\gamma_nt}.
    \label{eq:exactdecay}
\end{equation}
If the system is initially at rest at $t=0$, then $q_n(0)=Q_n$ and $\dot q_n(0)=0$.
Equation~\eqref{eq:envenergy} at $t=0$ then reduces to the usual modal elastic energy,
\begin{equation}
    \mathcal{E}_n(0)
    =
    \frac{\rho\omega_n^2Q_n^2}{2}
    =
    \frac{Gk_n^2Q_n^2}{2}.
    \label{eq:initial-energy}
\end{equation}

We therefore use
\begin{equation}
    \mathcal{E}_\mathrm{env}(t)
    =\sum_{n=1}^\infty \mathcal{E}_n(t)
    \label{eq:E_sum}
\end{equation}
as a measure of the relaxation process.
Unlike instantaneous displacement or stress, $\mathcal{E}_\mathrm{env}$ cannot exhibit a crossing solely as a result of differences in vibration phase.

On the other hand, the ordinary mechanical vibrational energy of the viscoelastic body is given by
\begin{equation}
    E_\mathrm{vib}(t)
    =
    \frac{1}{2}
    \int_0^L
    \left(
        \rho u_t^2
        +
        G u_x^2
    \right)\mathrm{d}x
    =
    \sum_{n=1}^{\infty}
    E_n^\mathrm{vib}(t),
    \label{eq:Evib}
\end{equation}
where
\begin{equation}
    E_n^\mathrm{vib}(t)
    =
    \frac{\rho}{2}
    \left(
        \dot{q}_n^2
        +
        \omega_n^2 q_n^2
    \right).
    \label{eq:Evib_mode}
\end{equation}
This is the mechanical energy in the usual sense, namely the sum of the kinetic energy and the elastic strain energy.

Using Eq.~\eqref{eq:pde} together with the fixed-end boundary conditions, we obtain
\begin{equation}
    \frac{\mathrm{d}E_\mathrm{vib}}{\mathrm{d}t}
    =
    -\eta
    \int_0^L
    u_{tx}^2\,\mathrm{d}x
    =
    -2\rho
    \sum_{n=1}^{\infty}
    \gamma_n\dot{q}_n^2
    \leq 0.
    \label{eq:Evib_decay}
\end{equation}
Thus, $E_\mathrm{vib}(t)$ itself decreases monotonically with time.

We emphasize that $\mathcal{E}_\mathrm{env}$ is not identical to $E_\mathrm{vib}$.
For each mode,
\begin{equation}
    \mathcal{E}_n
    =
    E_n^\mathrm{vib}
    +
    \rho\gamma_n q_n \dot{q}_n,
    \label{eq:env_vs_vib}
\end{equation}
and $\mathcal E_n$ may therefore be regarded as a relaxation measure obtained by removing the rapid oscillatory variation associated with the vibration phase from the mechanical energy.
In the following, we first use $\mathcal E_\mathrm{env}$ to clarify analytically the mechanism of the Mpemba crossing, and then show that the same reversal of relaxation ordering also occurs in the ordinary mechanical vibrational energy $E_\mathrm{vib}$.

\section{Mpemba relaxation in a two-mode system}\label{sec:Mpemba}
\subsection{Reversal of relaxation ordering in the envelope energy}
The essential mechanism is spectral in nature.
That is, the long-time relaxation is governed by the projection onto the slowest mode, as in other spectral realizations of the Mpemba effect~\cite{LuRaz2017,Klich2019,HayakawaTakada2026}.
This mechanism is already fully realized within an invariant two-mode subspace.
We assume that only the fundamental mode $n=1$ and the $m$th mode
($m>1$) are excited, and consider the following one-parameter family:
\begin{equation}
    Q_1=A\cos\theta,\quad
    Q_m=A\sin\theta,\quad
    0\leq\theta\leq\frac{\pi}{2},
    \label{eq:family}
\end{equation}
with $Q_n=0$ for all other modes ($n\neq 1,m$).
Since Eq.~\eqref{eq:pde} is linear and diagonal in the normal-mode
representation, imposing Eq.~\eqref{eq:family} together with the
initial-rest condition gives an exact two-mode solution rather than
a truncation approximation.

We define
\begin{equation}
    E_0=\frac{Gk_1^2A^2}{2}.
\end{equation}
From Eqs.~\eqref{eq:exactdecay}, \eqref{eq:initial-energy}, and \eqref{eq:E_sum}, we obtain
\begin{equation}
    \frac{\mathcal{E}_\mathrm{env}(t;\theta)}{E_0}
    =
    \cos^2\theta\,\mathrm{e}^{-2\gamma_1t}
    +m^2\sin^2\theta\,\mathrm{e}^{-2m^2\gamma_1t}.
    \label{eq:familyenergy}
\end{equation}
This expression contains the central result of the present analysis.
The initial excitation is given by
\begin{equation}
    \frac{\mathcal{E}_\mathrm{env}(0;\theta)}{E_0}
    =
    \cos^2\theta+m^2\sin^2\theta,
\end{equation}
which increases monotonically with $\theta$.
By contrast, the coefficient of the slowest relaxing contribution, $\cos^2\theta$, decreases monotonically with $\theta$.

We now choose two states H and C from this one-parameter family such that
\begin{equation}\label{eq:theta_HC}
    \theta_\mathrm{H}>\theta_\mathrm{C}.
\end{equation}
Here H and C correspond to \textit{Hotter} and \textit{Colder}, respectively, in the conventional Mpemba-effect terminology.
Denoting the modal envelope energies of states H and C by $\mathcal{E}_\mathrm{H}$ and $\mathcal{E}_\mathrm{C}$, respectively, their difference at $t=0$ is
\begin{equation}
    \frac{\mathcal{E}_\mathrm{H}(0)-\mathcal{E}_\mathrm{C}(0)}{E_0}
    =(m^2-1)
    \left(
    \sin^2\theta_\mathrm{H}-\sin^2\theta_\mathrm{C}
    \right)>0.
    \label{eq:initialorder}
\end{equation}
Thus, state H is initially farther from the relaxed state than state C.
In the long-time limit, however,
\begin{equation}
    \mathcal{E}_\mathrm{H}-\mathcal{E}_\mathrm{C}
    \sim E_0
    \left(
    \cos^2\theta_\mathrm{H}-\cos^2\theta_\mathrm{C}
    \right)
    \mathrm{e}^{-2\gamma_1t}<0.
    \label{eq:longtime}
\end{equation}
Therefore, the relaxation ordering must reverse at an intermediate time.

More interestingly, the crossing time is independent of the particular pair of states selected from this family.
Using $\sin^2\theta_\mathrm{H}-\sin^2\theta_\mathrm{C}=-\left(\cos^2\theta_\mathrm{H}-\cos^2\theta_\mathrm{C}\right)$, the condition
\begin{equation}
    \mathcal{E}_\mathrm{H}(t_\times)
    =
    \mathcal{E}_\mathrm{C}(t_\times)
\end{equation}
gives
\begin{equation}
    t_\times
    =
    \frac{\ln m^2}{2(m^2-1)\gamma_1}.
    \label{eq:crossing}
\end{equation}
Hence, every ordered pair in this one-parameter family satisfying Eq. \eqref{eq:theta_HC} crosses at the same dimensionless time.
This property originates from the fact that both the initial modal elastic energy and the Kelvin--Voigt decay rate are proportional to $n^2$.

\begin{figure}
    \centering
    \includegraphics[width=0.5\linewidth]{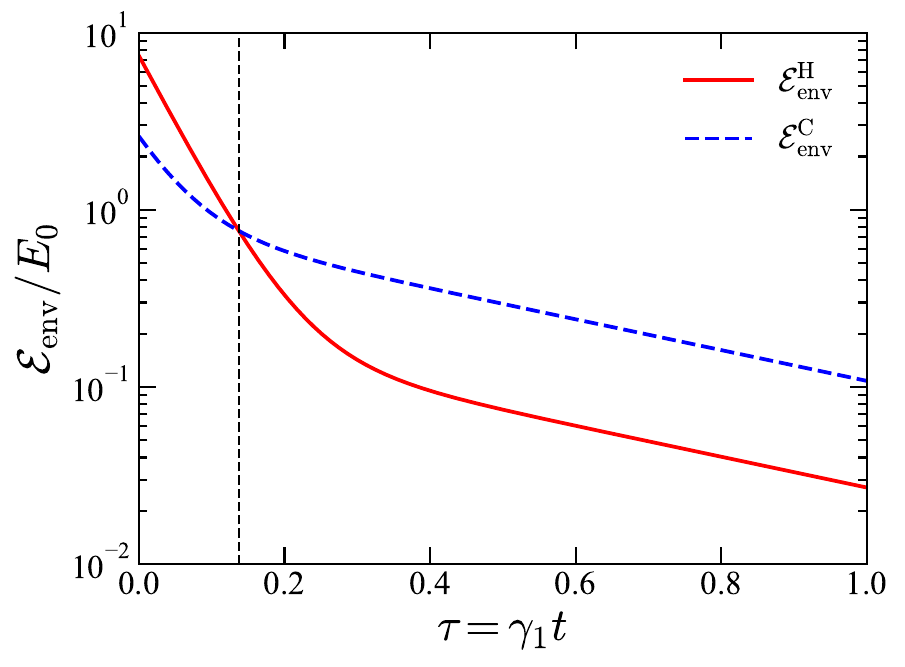}
    \caption{
    Relaxation of the envelope energy $\mathcal{E}_\mathrm{env}$ for $m=3$, $\theta_\mathrm{H}=\tan^{-1}2$, and $\theta_\mathrm{C}=\tan^{-1}(1/2)$.
    The solid and dashed lines represent $\mathcal{E}_\mathrm{env}$ for states H and C, respectively.
    The vertical line indicates the crossing time $\tau_\times$ of $\mathcal{E}_\mathrm{env}$.
    }
    \label{fig:crossing}
\end{figure}

Figure~\ref{fig:crossing} shows the time evolution of the two states for the choice
\begin{equation}
    m=3,\quad
    \theta_\mathrm{H}=\tan^{-1}2,\quad
    \theta_\mathrm{C}=\tan^{-1}\frac{1}{2}.
\end{equation}
From Eq.~\eqref{eq:familyenergy}, we explicitly obtain
\begin{align}
    \frac{\mathcal{E}_\mathrm{H}}{E_0}
    &=\frac{1}{5}\mathrm{e}^{-2\tau}
      +\frac{36}{5}\mathrm{e}^{-18\tau},\\
    \frac{\mathcal{E}_\mathrm{C}}{E_0}
    &=\frac{4}{5}\mathrm{e}^{-2\tau}
      +\frac{9}{5}\mathrm{e}^{-18\tau},
\end{align}
where
\begin{equation}
    \tau:=\gamma_1t
\end{equation}
is the dimensionless time.
At the initial time,
\begin{equation}
    \frac{\mathcal{E}_\mathrm{H}(0)}{E_0}
    =\frac{37}{5},\quad
    \frac{\mathcal{E}_\mathrm{C}(0)}{E_0}
    = \frac{13}{5},
\end{equation}
whereas the two curves cross, according to Eq.~\eqref{eq:crossing}, at
\begin{equation}
    \tau_\times
    =
    \frac{\ln9}{16}
    \simeq0.1373.
\end{equation}
After the crossing, the envelope energy of state H becomes smaller than that of state C because the weight of the slowest mode in H is only one quarter of that in C.

\subsection{Reversal of relaxation ordering in the mechanical vibrational energy}
We next show that the reversal of relaxation ordering discussed above is not an artifact of the definition of the envelope energy $\mathcal{E}_\mathrm{env}$ introduced for analytical convenience.
The ordinary mechanical vibrational energy $E_n^\mathrm{vib}(t)$, defined in Eqs \eqref{eq:Evib} and \eqref{eq:Evib_mode}, is the mechanical energy in the usual sense, namely the sum of the kinetic and elastic strain energies.
For the present initial condition, the modal contribution can be written as
\begin{equation}
    E_n^\mathrm{vib}(t)
    =
    E_n^\mathrm{vib}(0)F_n(t),
    \label{eq:Fn_def}
\end{equation}
where
\begin{equation}
    F_n(t)
    =
    \mathrm{e}^{-2\gamma_n t}
    \left\{
        \left[
            \cos(\Omega_n t)
            +
            \frac{\gamma_n}{\Omega_n}
            \sin(\Omega_n t)
        \right]^2
        +
        \frac{\omega_n^2}{\Omega_n^2}
        \sin^2(\Omega_n t)
    \right\},
    \label{eq:Fn}
\end{equation}
with $F_n(0)=1$.

Therefore, for two states satisfying $\theta_\mathrm{H}>\theta_\mathrm{C}$, the difference in their mechanical vibrational energies can be factorized as
\begin{equation}
    \frac{
        E_\mathrm{vib}^\mathrm{H}(t)
        -
        E_\mathrm{vib}^\mathrm{C}(t)
    }{E_0}
    =
    \left(
        \cos^2\theta_\mathrm{H}
        -
        \cos^2\theta_\mathrm{C}
    \right)
    \left[
        F_1(t)-m^2F_m(t)
    \right].
    \label{eq:Evib_difference}
\end{equation}
At $t=0$, $E_\mathrm{vib}^\mathrm{H}(0)>E_\mathrm{vib}^\mathrm{C}(0)$, whereas in the long-time regime, $E_\mathrm{vib}^\mathrm{H}(t)<E_\mathrm{vib}^\mathrm{C}(t)$.
Thus, a reversal of relaxation ordering necessarily occurs also in the ordinary mechanical vibrational energy.
Unlike the case of the envelope energy, however, the crossing time depends on the ratio between the oscillation frequency and the damping rate.

\begin{figure}
    \centering
    \includegraphics[width=0.75\linewidth]{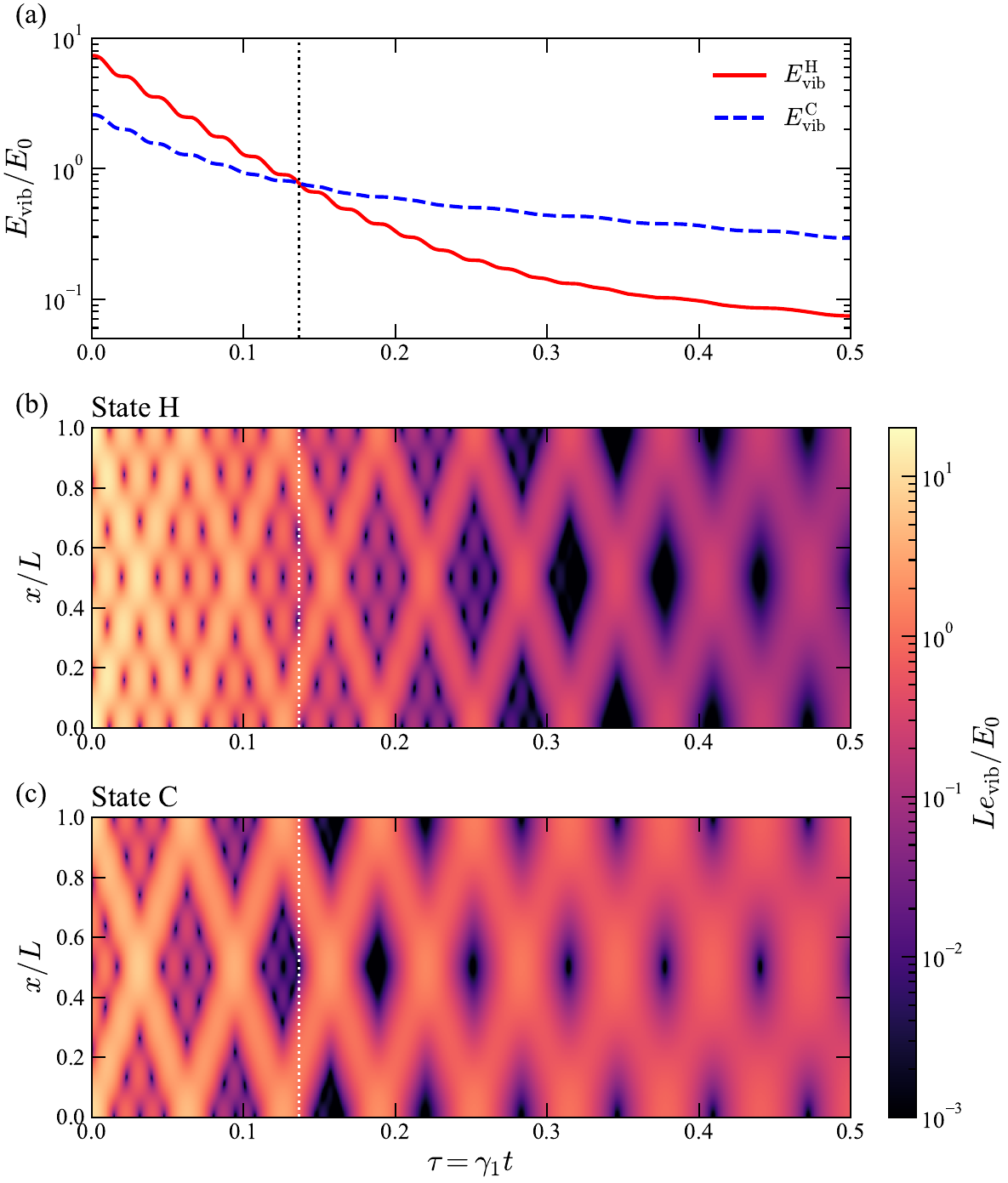}
    \caption{
    (a) Relaxation of the ordinary mechanical vibrational energy $E_\mathrm{vib}$ for $m=3$, $\theta_\mathrm{H}=\tan^{-1}2$, and $\theta_\mathrm{C}=\tan^{-1}(1/2)$.
    The solid and dashed lines represent $E_\mathrm{vib}$ for states H and C, respectively.
    The vertical line indicates the crossing time $\tau_\times$ of $\mathcal{E}_\mathrm{env}$.
    Here, we set $R=\omega_1/\gamma_1=50$.
    (b), (c) Spatiotemporal evolution of the local mechanical vibrational energy density for states H and C, respectively.
    The color represents the dimensionless local energy density $L e_\mathrm{vib}(x,t)/E_0$ on a logarithmic scale.
    }
    \label{fig:E_vib}
\end{figure}

The time evolution of the mechanical vibrational energy is shown in Fig.~\ref{fig:E_vib}(a).
Here, we set
\begin{equation}
    R
    := \frac{\omega_1}{\gamma_1}
    = 50.
\end{equation}
The mechanical vibrational energies cross at a time very close to the exact envelope-energy crossing time $\tau_\times^\mathrm{env}$.
As can also be seen in Fig.~\ref{fig:E_vib} (a), $E_\mathrm{vib}$ exhibits small modulations associated with the vibration phase, but decays approximately along the same relaxation envelope as $\mathcal E_\mathrm{env}$.
The relaxation ordering between states H and C is therefore reversed also in the actual mechanical energy.
This demonstrates that the Mpemba-type crossing found here is not artificially generated by introducing the auxiliary relaxation measure $\mathcal E_\mathrm{env}$.

Figures~\ref{fig:E_vib}(b) and (c) show the corresponding spatiotemporal distributions of the local mechanical vibrational energy density,
\begin{equation}
    e_\mathrm{vib}(x,t)
    =
    \frac{1}{2}
    \left[
        \rho u_t^2(x,t)
        +
        G u_x^2(x,t)
    \right].
    \label{eq:local_Evib}
\end{equation}
The plotted quantity is $L e_\mathrm{vib}(x,t)/E_0$, so that its spatial integral over $x/L$ yields the normalized total energy $E_\mathrm{vib}(t)/E_0$.

The difference between the two relaxation processes is also apparent at the level of the spatial energy distribution.
State H, shown in Fig.~\ref{fig:E_vib}(b), contains a larger contribution from the higher mode at the initial time.
Accordingly, its energy-density profile exhibits relatively fine spatial structures and rapid temporal oscillations.
Since the damping rate increases with the mode number, these higher-mode contributions are strongly attenuated at early times.
As a result, the initially larger mechanical energy of state H decreases rapidly.

By contrast, state C in Fig.~\ref{fig:E_vib}(c) contains a larger relative contribution from the fundamental mode.
Its energy-density distribution therefore has a broader spatial structure and decays more slowly.
Around the crossing time indicated by the vertical dotted line, the rapidly relaxing higher-mode contribution in state H has already been substantially suppressed, whereas the slowly relaxing fundamental-mode contribution in state C remains appreciable.
This difference in modal composition provides a direct spatiotemporal picture of the reversal of the relaxation ordering observed in Fig.~\ref{fig:E_vib}(a).

\section{Strong vibrational Mpemba limit}\label{sec:strong_Mpemba}
The endpoint $\theta_\mathrm{H}=\pi/2$ is particularly noteworthy.
In this state, the slowest fundamental mode is completely absent, and
\begin{equation}
    \mathcal{E}_\mathrm{H}(t)
    =
    m^2E_0\,\mathrm{e}^{-2m^2\gamma_1t}.
\end{equation}
By contrast, any reference state satisfying $\theta_\mathrm{C}<\pi/2$ retains a contribution from the fundamental mode proportional to $\mathrm{e}^{-2\gamma_1t}$.

Therefore, when the projection onto the slowest mode vanishes, the energy decay rate governing the long-time behavior changes from
\begin{equation}
    2\gamma_1
    \Rightarrow
    2m^2\gamma_1.
\end{equation}
This is a direct vibrational analogue of the strong Mpemba mechanism~\cite{Klich2019,Kumar2020,Biswas2020,Zhang2025}, in which anomalously rapid relaxation is achieved by suppressing the slowest relaxation channel.
This result is also consistent with the more general observation that, when the nominally slowest mode is absent or effectively inactive, anomalous relaxation can be governed entirely by faster modes~\cite{Chatterjee2023}.

The same conclusion holds for the ordinary mechanical vibrational energy $E_\mathrm{vib}$.
For $\theta_\mathrm{H}=\pi/2$, the projection onto the fundamental mode is exactly zero, so that the long-time decay of $E_\mathrm{vib}^\mathrm{H}$ is also governed by the $m$th mode.
By contrast, a reference state with $\theta_\mathrm{C}<\pi/2$ retains the fundamental mode, and its long-time relaxation is governed by an envelope proportional to $\mathrm{e}^{-2\gamma_1t}$.
Hence, the strong Mpemba mechanism does not depend on the definition of the envelope energy.

\section{Discussion and conclusion}\label{sec:conclusion}
The crossing obtained here does not arise simply because one state starts with a larger overall amplitude.
If two initial states differ only by an overall scale factor and satisfy
\begin{equation}
    Q_n^\mathrm{H}=aQ_n^\mathrm{C}
\end{equation}
for all $n$, the linearity of the system implies
\begin{equation}
    \mathcal E_\mathrm{H}(t)
    =
    a^2\mathcal E_\mathrm{C}(t)
\end{equation}
at all times, and no crossing can occur.
The essential requirement is a redistribution of the initial excitation from slowly relaxing modes to more rapidly decaying modes.

This also explains why mode-dependent decay rates are necessary for the present mechanism.
If the Kelvin--Voigt viscous term is replaced by a spatially uniform drag term~\cite{Woodhouse1998,Adhikari2006},
\begin{equation}
    \rho u_{tt}
    + \zeta u_t
    = G u_{xx},
\end{equation}
the amplitude decay rate is identical for every mode,
\begin{equation}
    \gamma
    =
    \frac{\zeta}{2\rho}.
\end{equation}
In this case, the envelope energy of every mode acquires the same time-dependent factor, and the ordering between two states cannot reverse during the relaxation.
More generally, the present mechanism is not specific to continua, but applies to linear mechanical systems composed of multiple independent normal modes with different decay rates.

The family of initial states introduced in Eq.~\eqref{eq:family} can, in principle, be prepared by mode-selective excitation.
For example, one may apply a finite-duration external force whose projections onto $\phi_1$ and $\phi_m$ are controlled so as to set the mixing angle $\theta$, and then remove the force at $t=0$.
Similar initial states may also be prepared using spatially patterned forcing or resonant excitation.
In either case, the linear operator governing the subsequent free relaxation is exactly the same for states H and C, and the anomalous relaxation ordering originates solely from the difference in their initial modal weights.

The present results place a standard damped elastic medium within a broader spectral picture of Mpemba-type relaxation acceleration.
We first introduced the positive modal envelope energy $\mathcal E_\mathrm{env}$ in order to eliminate trivial crossings caused by differences in vibration phase.
This quantity makes it possible to identify the mechanism of Mpemba relaxation in the Kelvin--Voigt system analytically, and the crossing time can be obtained exactly within the two-mode subspace.

Importantly, however, the reversal of relaxation ordering found here is not dependent on this auxiliary definition of the envelope energy.
The ordinary mechanical vibrational energy decreases monotonically with time, yet still exhibits a reversal of relaxation ordering between the same two initial states.
Therefore, the Mpemba-type crossing found here is not an artifact of vibration phase or of a specially constructed relaxation measure, but a physical relaxation phenomenon that also appears in the ordinary mechanical energy.

In the minimal Kelvin--Voigt model, $\gamma_n\propto n^2$.
Thus, by redistributing the initial excitation from the fundamental mode to higher modes, one can simultaneously increase the total initial vibrational energy and reduce the projection onto the slowest relaxing mode.
The coexistence of these two properties is the essential ingredient of Mpemba-type relaxation in the present linear mechanical system.

Furthermore, when the projection onto the slowest mode vanishes completely, the decay rate governing the long-time relaxation switches to that of a faster mode for both the envelope energy and the ordinary mechanical vibrational energy.
The present system therefore provides a simple realization of the strong Mpemba effect in a linear vibrational setting.

The mechanism demonstrated here is not specific to the Kelvin--Voigt continuum, but should apply to a broad class of linear mechanical systems composed of multiple normal modes with different decay rates~\cite{CaugheyOKelly1965,Woodhouse1998,Adhikari2006}.
Possible extensions include multidimensional elastic structures, discrete mass--spring--damper systems, and open elastodynamic systems with mode-dependent damping caused by wave radiation.

\begin{acknowledgments}
This work is partially supported by the Grant-in-Aid of MEXT for Scientific Research (Grant Nos.~JP24K06974, JP24K07193, JP24KJ0110, JP25K01063, and JP26K06964).

\end{acknowledgments}

\appendix

\section{Derivation of the modal envelope energy}
\label{sec:derivation}
In this Appendix, we briefly derive the modal envelope energy.

Using the transformation in Eq.~\eqref{eq:transform}, we obtain
\begin{equation}
    \dot{q}_n
    =
    \mathrm{e}^{-\gamma_n t}
    \left(
        \dot{y}_n
        -
        \gamma_n y_n
    \right)
\end{equation}
and
\begin{equation}
    \ddot q_n
    =
    \mathrm{e}^{-\gamma_n t}
    \left(
        \ddot{y}_n
        -
        2\gamma_n \dot{y}_n
        +
        \gamma_n^2 y_n
    \right).
\end{equation}
Substituting these expressions into the equation of motion, Eq.~\eqref{eq:mode}, eliminates the first-order time-derivative term and gives
\begin{equation}
    \ddot{y}_n
    +
    \Omega_n^2 y_n
    = 0,
    \label{eq:app_yosc}
\end{equation}
where
\begin{equation}
    \Omega_n^2
    = \omega_n^2
    - \gamma_n^2.
\end{equation}
Thus, under the underdamped condition $\gamma_n<\omega_n$, $y_n$ obeys the equation of an undamped harmonic oscillator with angular frequency $\Omega_n$.

For the undamped oscillator $y_n$, the quantity
\begin{equation}
    E_{y,n}
    =
    \frac{\rho}{2}
    \left(
        \dot y_n^2
        +
        \Omega_n^2 y_n^2
    \right)
    \label{eq:app_Ey}
\end{equation}
is conserved.
On the other hand, Eq.~\eqref{eq:transform} gives
\begin{equation}
    y_n
    = \mathrm{e}^{\gamma_n t} q_n,\quad
    \dot y_n
    = \mathrm{e}^{\gamma_n t}
    \left(\dot q_n + \gamma_n q_n\right).
\end{equation}
Therefore, Eq.~\eqref{eq:app_Ey} can be rewritten as
\begin{equation}
    E_{y,n}
    =
    \mathrm{e}^{2\gamma_n t}
    \frac{\rho}{2}
    \left[
        \left(
            \dot q_n
            +
            \gamma_n q_n
        \right)^2
        +
        \Omega_n^2 q_n^2
    \right].
\end{equation}

We therefore define
\begin{equation}
    \mathcal{E}_n(t)
    =
    \frac{\rho}{2}
    \left[
        \left(
            \dot q_n
            +
            \gamma_n q_n
        \right)^2
        +
        \Omega_n^2 q_n^2
    \right]
    \label{eq:app_envenergy}
\end{equation}
as the modal envelope energy.
Since Eq.~\eqref{eq:app_Ey} is a conserved quantity,
\begin{equation}
    \mathcal{E}_n(t)
    =
    E_{y,n}\,
    \mathrm{e}^{-2\gamma_n t},
\end{equation}
and hence Eq.~\eqref{eq:exactdecay} holds exactly.
Under the underdamped condition, $\mathcal{E}_n(t)$ is positive and directly represents the monotonic exponential relaxation of each mode.



\end{document}